\documentclass[12pt]{JHEP3}
\usepackage{amsfonts} 
\usepackage{amsbsy} 
\usepackage{amssymb} 
\usepackage{cite}

\newcommand{\C}[1]{{\mathcal #1}}
\newcommand{\BS}[1]{{\boldsymbol{ #1}}}

\newcommand{\beq}{\begin{equation}}
\newcommand{\eeq}{\end{equation}}
\newcommand{\bea}{\begin{eqnarray}}
\newcommand{\eea}{\end{eqnarray}}

\newcommand{\Tr}{{\hbox{Tr}\,}}
\newcommand{\comm}[2]{\left[#1,#2\right]}
\newcommand{\absval}[1]{{\left\vert#1\right\vert}}

\newcommand{\half}{{1\over 2}}
\newcommand{\threehalves}{{3\over 2}}

\newcommand{\quarter}{{1\over 4}}

\newcommand{\nn}{\nonumber}
\newcommand{\Xperp}{\overline{X}}

\newcommand{\Prodroot}[1]{\prod_{\positive{#1}}}
\newcommand{\Sumroot}[1]{\sum_{\positive{#1}}}
\newcommand{\weyldet}[1]{\Prodroot{\alpha} \left( #1 \cdot \alpha
\right)^2}
\newcommand{\positive}[1]{#1\in\Delta_+}

\def\Pf{\C P}
\def\const{ \hbox{\it{const}}}
\def\Qbar{\overline{Q}}
\def\Qperp{{Q^\perp}}

\title{Polyakov Lines in Yang-Mills  Matrix Models}

\author{Peter Austing\\ Department of Physics, University of Oxford \\
Theoretical Physics,\\
1 Keble Road,\\
 Oxford OX1 3NP, UK\\
E-mail: \email{p.austing@physics.ox.ac.uk}}

\author{Graziano Vernizzi\\ Service de Physique Th\'eorique,
CEA/DSM/SPhT Saclay,\\ Unit\'e de recherche associ\'ee au
CNRS\\F-91191 Gif-sur-Yvette C\'edex, France\\
E-mail: \email{vernizzi@spht.saclay.cea.fr}}

\author{John F. Wheater\\ Department of Physics, University of Oxford \\
Theoretical Physics,\\
1 Keble Road,\\
 Oxford OX1 3NP, UK\\
E-mail: \email{j.wheater@physics.ox.ac.uk}}

\preprint{hep-th/0309026\\ OUTP/03-23P, SPhT-T03/129}
\abstract{We study the Polyakov line in Yang-Mills matrix models,
which include the IKKT model of IIB string theory. For the gauge
group  $SU(2)$ we give the exact formulae in the form of integral
representations which are convenient for finding the asymptotic 
behaviour. For the $SU(N)$ bosonic models we prove
upper bounds which decay  as a power law  at  large momentum $p$.
We argue that these capture the full asymptotic behaviour.
 We also indicate how to
extend the results to some correlation functions of Polyakov lines.}


\begin{document}

\section{Introduction}

The proposal \cite{Ishibashi:1997xs} for a matrix model describing IIB
string theory has triggered considerable interest in Yang-Mills matrix
models. The models are defined in section \ref{sec:defn}, and it
 is the $D=10$ dimensional maximally
supersymmetric version which is proposed as the model of string theory.
 When the gauge group is $SU(2)$ it has been known for
some time how to compute the supersymmetric integrals
\cite{Savvidy:1985gi,Smilga:1986jg,Smilga:1986nt,Yi:1997eg,Sethi:1997pa,Suyama:1998ig}.
It was believed that the bosonic integrals (in which the fermionic degrees
of freedom are omitted)  would diverge because of
the flat directions in which the matrices all commute, but the authors of
\cite{Krauth:1998yu} computed them analytically for $SU(2)$ and
numerically for some other groups, and realised they can converge as long
as $D$ is big enough. We established analytically the convergence
criteria for the bosonic $SU(N)$ integrals in \cite{Austing:2001bd}
and for the other gauge groups and supersymmetric integrals in
\cite{Austing:2001pk}, confirming that the partition function does
exist in the bosonic case when $D$ is big enough (except for $SU(2)$
and $SU(3)$, $D$ need only be bigger than $2$), and that the
supersymmetric models exist when $D=4$, $6$ and $10$. For further details,
see also \cite{Austing:2001ib}.

For  $SU(N\!>\!2)$ it is not known how to compute the integrals
exactly, but the authors of \cite{Moore:1998et} used the supersymmetry
to deform 
the partition function  into a cohomological theory in
which the integrals can be done. Although the relation of this model
to the original Yang-Mills model remains
unproven, the resulting numbers have been confirmed by very careful
numerical 
calculation for small values of $N$ \cite{Krauth:1998xh,Krauth:1998yu}, and support a conjecture \cite{Green:1998yf} based on D brane dynamics.
The method has also been extended to the other simple groups, and checked numerically for
small rank \cite{Krauth:2000bv,Staudacher:2000gx} and see \cite{Pestun:2002rr}. Unfortunately, it
cannot be extended to observables such as the Polyakov line 
which break supersymmetry. 

Another important line of research has been into overcoming the
considerable difficulties in obtaining numerical results for larger
values of $N$ in order to gain understanding of the large $N$
limit. Various simulations of the bosonic $SU(N)$ model have been
reported
\cite{Hotta:1998en,Ambjorn:2000dj,Horata:2000ft,Kitsunezaki:2000ff,Kitsunezaki:2001bt},
and the Polyakov line and Wilson loop have been studied up to $N=768$ in \cite{Anagnostopoulos:2001cb}.
For the supersymmetric model with $D=4$, the Pfaffian is positive, and
these models have been studied up to $N=48$
\cite{Ambjorn:2000bf,Burda:2000mn,Ambjorn:2000dj}, and
with particular reference to Polyakov lines and Wilson loops \cite{Ambjorn:2001xs}. For $D=6$ and $D=10$, the
situation is much more difficult since the Pfaffian is
complex. Various ways of dealing with this have been tried
\cite{Ambjorn:2000dx,Nishimura:2000wf,Nishimura:2000ds,Vernizzi:2002mu}.
 Meanwhile, supersymmetric random surface, and geometrical approaches
have been investigated in \cite{Bialas:2000gf,Burda:2002vy},
as have mean field theory  and  $1/D$ expansions in \cite{Aoki:1998vn,Oda:2000im,Sugino:2001,Nishimura:2001sx,Kawai:2002jk,Kawai:2002ub}.

The purpose of this paper is to obtain rigorous information about
 the large $p$ behaviour of the Polyakov line for $SU(N)$ with finite $N$. 
The scene is set in section 2. In  section 3 we
 derive the exact results for the bosonic and
 supersymmetric $SU(2)$ models; this is not new but our derivation 
gives an integral representation which enables the asymptotic behaviour 
to be calculated very easily. In section 4  we derive upper
 bounds on the large $p$ behaviour for the Polyakov line in the bosonic
theory. These decay polynomially, and as we shall indicate, there is good 
reason to believe that the upper bounds capture the true behaviour. 
We also indicate how to extend the results to some correlation
 functions of Polyakov lines. Section 5 contains a brief discussion.

\section{Definition of the models}
\label{sec:defn}
The  Yang-Mills matrix integral partition function, 
which is obtained by dimensionally reducing the Euclidean SSYM
action from $D$ down to zero dimensions,  is given by
\beq \C Z_{D,G} = \int 
\prod_{\mu=1}^D dX_\mu 
 \prod_{\alpha=1}^{\C{N}} d \psi_{\alpha}
\exp \left( \quarter \sum_{\mu , \nu}
\Tr  \comm{X_\mu}{X_\nu}^2 + \half \Tr \psi_{\alpha} [ \Gamma_{\alpha
\beta}^{\mu} X_{\mu}, \psi_{\beta}] \right), \label{P1}
\eeq
where we adopt the summation convention for repeated indices.
The traceless hermitian matrix fields $X_\mu$ and $\psi_\alpha$ (respectively bosonic and
fermionic) are in the Lie algebra $\C G$ of the (compact semi-simple)
 gauge group $G$ and can be written
\beq 
X_\mu=\sum_{a=1}^g X_\mu^a t^a \; , \;\;\;\; \psi_{\alpha} = \sum_{a=1}^g
\psi_{\alpha}^a t^a,
\label{P2}
\eeq
where $\{t^a, a=1,\ldots ,g\}$ are the generators in the fundamental
representation. The $\Gamma^\mu_{\alpha \beta}$ are ordinary gamma
matrices for $D$ Euclidean dimensions. The model possesses a gauge symmetry
\beq X_\mu\to U^\dagger X_\mu U,\quad\psi_\alpha\to U^\dagger \psi_\alpha U,\qquad U\in G. \label{gaugetr}\eeq
and an $SO(D)$ symmetry inherited from the original $D$-dimensional
Euclidean symmetry of the SSYM. 
 Although the  motivation discussed above leads to a study
of the $D=10$, $SU(N)$ supersymmetric integral, it is useful and illuminating to
study several different versions  of the model. Firstly by suppressing
the fermions we get the bosonic integrals which we will denote by $\C N=0$
(ie there are no super-charges)  \cite{Krauth:1998yu}. 
Secondly the supersymmetric integrals can be written
 for $D=3$, 4, 6, and 10, having
$\C N=2(D-2)$ super-charges.
In principle one can integrate out the fermions to obtain
\beq
\label{1.3}
\C Z_{D,G} =\int 
\prod_{\mu=1}^D dX_\mu \,
\Pf_{D,G} (X_\mu )
\exp \left( -S_D(X) \right),
\eeq
where
\beq
S_D(X)=-\quarter \sum_{\mu ,\nu=1}^D
\Tr \comm{X_\mu}{X_\nu}^2
\eeq
and the Pfaffian $\Pf_{D,G}$ is a homogeneous polynomial of degree
$\half \C N g$. 
In this paper we will be concerned with Polyakov line
   correlation functions of the form
\bea\label{correl}
L(P)&=&<\Tr\exp(i P^\sigma X_\sigma )>\nn\\
&=& \int 
\prod_{\mu=1}^D dX_\mu \, \Tr\exp(i P^\sigma X_\sigma ) \,
\Pf_{D,G} (X_\mu )
\exp \left( \quarter \sum_{\mu ,\nu}
\Tr \comm{X_\mu}{X_\nu}^2 \right).
\eea
Using the SO(D) symmetry we can always consider $P^\sigma=(p,0,\ldots)$
and denote the corresponding Polyakov line by $L(p)$.

It is convenient to express the Lie algebra $\C G$ using the Cartan-Weyl basis
\beq
\label{B11}
\{ H^i, E^\alpha \},
\eeq
where $i$ runs from $1$ to the rank $r$ and $ \alpha $ denotes
a root.
In this basis
\beq
\comm{H^i}{H^j}=0\, , \;\;\;\; \comm{H^i}{E^\alpha}=\alpha^i E^\alpha
\eeq
and
\beq
\begin{array}{llll}
\comm{E^\alpha}{E^\beta}&=& N_{\alpha
\beta}E^{\alpha+\beta}\;\;\;\;\;\; & \hbox{if } \alpha + \beta \hbox{
is a root}\\
& =&2
\absval{\alpha}^{-2}\, \alpha \cdot H& \hbox{if } \alpha = -\beta \\
&=&0& \hbox{otherwise.}
\end{array}
\eeq
Here $E^{-\alpha} = (E^{\alpha})^{\dagger}$, and the normalisation is
chosen  such that 
\beq
\label{innerprod}
\Tr H^i H^j = \delta^{ij} \, , \;\;\; \Tr E^\alpha E^{\beta} =
2  \absval{\alpha}^{-2}\, \delta^{\alpha+ \beta},\qquad \Tr H^iE^\alpha=0.
\eeq
 We
can expand any matrix in the Lie Algebra  as
\beq
M = \widetilde M\cdot H +\sum_\alpha \overline {M}^\alpha E^\alpha 
\eeq
with 
$\overline {M}^{-\alpha} = (\overline {M}^\alpha)^*$
 (but for readability's sake we will only use this expansion when
necessary).
We will denote the set of all roots by $\Delta$, the set of positive
roots by $\Delta_+$, and the set of simple positive roots by $\Delta_+^*$;
in addition we will denote a simple root by $s$, retaining $\alpha$ for 
a generic root. The dual weights $\omega_s$ satisfy
\beq\label{2.5}  2{s\cdot \omega_{s'} \over \absval{s}^2}=\delta_{ss'}.\eeq
The above definitions (\ref{B11} - \ref{2.5}) fix the normalisation
of the long root to be $\sqrt{2}$. In the case of $su(N)$, 
all roots are of equal magnitude $\sqrt{2}$.

Since the integrand and measure are gauge invariant, we can always make
a gauge transformation (\ref{gaugetr})
to move $X_D$ into the Cartan subalgebra
\beq
X_D = \lambda_D\cdot H
\eeq
and reduce the integral over $X_D$ to an
integral over its Cartan modes \cite{Krauth:2000bv}
\beq \prod_{a=1}^{g} dX_D^a \rightarrow \Omega_G\left( \prod_{i=1}^l d\lambda_D^i
\right) \Delta^2_G(\lambda_D),\eeq
where $\Omega_G$ is the volume of $G$ and the Weyl measure, given by
\beq
\label{vandermonde}
\Delta^2_G(\lambda_D) =\weyldet{\lambda_D},
\eeq
is the generalisation from $SU(N)$ of the Vandermonde determinant factors.

We will consider mainly the integral 
(\ref{correl}) without fermions so that $\C N
=0$ and there is no Pfaffian.
In the Cartan representation the expression (\ref{correl}) takes the form
\bea\label{LcartanA} L(p)&=&\int 
 \left( \prod_{i=1}^r d\lambda_D^i\right)
 \Delta^2_G(\lambda_D)\prod_{k=1}^{D-1} dX_k \, \Tr\exp(i p \lambda_D\cdot H )\nn\\
&&\exp \left(-\sum_{\alpha >0}
\frac{(\lambda_D \cdot \alpha)^2}{\absval{\alpha}^2}
\sum_{k=1}^{D-1}
\absval{ \Xperp_k^\alpha}^2+ \quarter \sum_{j,k=1}^{D-1}
\Tr \comm{X_j}{X_k}^2 \right)\eea
and since the weights in a given representation correspond to the
eigenvalues of the Cartan generators, we have
\beq
\Tr \exp(i p \lambda_D\cdot H ) = \sum_h \exp(i p \lambda_D\cdot h ),
\eeq
where $\{ h \}$ are the weights.
In the case of $su(N)$, every weight in the fundamental representation
is a Weyl transformation of every other weight, so we can use Weyl
invariance to write
\beq\label{2.12}
L(p)= N \int d \lambda_D  \Delta^2_G(\lambda_D) dX_k
\,  \exp \left(i p \lambda_D\cdot \omega_1-\sum_{\alpha >0}
\frac{(\lambda_D \cdot \alpha)^2}{\absval{\alpha}^2}
\sum_{k=1}^{D-1}
\absval{ \Xperp_k^\alpha}^2-S_{D-1} \right),
\eeq
since $\omega_1$ is the first weight in the fundamental representation.
\label{prelim}

\section{Exact calculations for $\BS{su(2)}$}
In the simplest case of $su(2)$ it is possible to reduce $L(p)$
to a simple integral representation. This is because
there are few enough fields to make good use of the rotation
symmetry.\footnote{The authors of \cite{Cicuta:2001ke} also managed to extend
this to the integrals over the $3 \! \times \! 3$ real symmetric matrices.}
Several authors have used exact
calculations of the $su(2)$ partition function in various contexts
\cite{Savvidy:1985gi,Smilga:1986jg,Smilga:1986nt,Yi:1997eg,Sethi:1997pa,Suyama:1998ig}, and
integral representations of the eigenvalue density have been given in
\cite{Krauth:1999rc}. In principle one only need take the Fourier
transform of the densities given in \cite{Krauth:1999rc} to obtain the
Polyakov line but  it is hard to extract the asymptotic
behaviour from these representations. The following calculations
amount to an alternative representation for the eigenvalue densities
from which the asymptotic behaviour of $L(p)$ can be easily
determined.

The definitions in section \ref{prelim} fix the normalisation of the
generators, but to avoid any ambiguity in the definition (\ref{correl})
 we give them explicitly
here;
\beq
\sigma_1={1 \over \sqrt{2}} \left( \begin{array}{cc} 0&1\\1&0
\end{array} \right) \, , \;\;\; \sigma_2={1 \over \sqrt{2}} \left( \begin{array}{cc} 0&-i\\i&0
\end{array} \right) \, , \;\;\; \sigma_3={1 \over \sqrt{2}} \left( \begin{array}{cc} 1&0\\0&-1
\end{array} \right)
\eeq
so that $H=\sigma_3$ and $E^{\pm 1}= \half (\sigma_1 \pm i \sigma_2)$.
\subsection{Bosonic}
There is only one positive root and (\ref{2.12}) simplifies
to 
\beq\label{Lsu2A} L(p)=2\int  d\lambda_D\,
\lambda_D^2
\prod_{k=1}^{D-1} dX_k \, \exp(i p \lambda_D/\sqrt{2} ) \,
\exp \left(- \lambda_D^2 \sum_{k=1}^{D-1}
\absval{ \Xperp_k^1}^2-S_{D-1} \right).\eeq
Setting
\beq
X_k= u_k \sigma_1+ v_k \sigma_2  + \xi_k \sigma_3, \quad k=1 \ldots d,
\eeq
where $d=D-1$, we get
\beq
L(p)=- 2 {d^2 \over dp^2} K(p),
\eeq
with
\beq
K(p)=\int d \lambda e^{ip\lambda}\prod_{k=1}^{D-1} d \xi_k du_k dv_k
\exp\left( -\half \sum_{i,j=1}^d \xi_i U_{ij} \xi_j -\quarter V-{\lambda^2 \over 2}W   \right),
\eeq
where
\bea
U_{ij}&=& (u^2+v^2) \delta_{ij}
-(u_i u_j +v_i v_j)\\
V &=& u^2 v^2 -(u \cdot v)^2\\
W&=& u^2+v^2.
\eea
Integrating out the $d$-dimensional vector $\xi$ gives, up to a
constant factor which we will fix later,
\beq
K(p)=\int d\lambda du \, dv \, (\det U)^{-\half} \exp(ip\lambda/\sqrt{2}
-V/4-\lambda^2 W/2)
\eeq
which, noting that the determinant is $\det U=W^{d-2}V$, we rewrite as
\beq
\int dA dB d\lambda \, {\exp(-\lambda^2 A/2 +ip\lambda/\sqrt{2}-B/4) \over A^{d-2
\over 2} B^\half} J(A,B),
\eeq
where
\bea
J&=&\int du \, dv \, \delta(A-W) \delta(B-V)\\
\label{2.23}&=& \const \, B^{d-3 \over 2} \, \theta\left( {A^2 \over 4}-B \right),
\eea
as one can check by using the rotation invariance.
Then integrating out $\lambda$, and scaling $A$ by $p^2$, we obtain
\beq
K(p)={(D-2) \over \Gamma\left({D \over 4}-1 \right)} \, p^{4-D} \int_0^\infty dB \, B^{d-4 \over 2} e^{-B/4}
\int_{2 \sqrt{B} \over p^2}^\infty dA \, A^{-{d-1 \over 2}} e^{-{1 \over 4A}},
\eeq
where we have fixed the constant factor by requiring $L(0)=1$.
Differentiating twice with respect to $p$, we
obtain terms which vanish like a polynomial from differentiating
$p^{4-D}$ and terms which vanish exponentially from differentiating
the integral. Then we find
\beq
L(p) \sim - 2^{D-1} \sqrt{\pi}{\Gamma(D-1) \over \Gamma\left({D \over 4}-1 \right)} \, p^{2-D}\label{su2upper}
\eeq
as $p \rightarrow \infty$ for $D>4$.\footnote{For $su(2)$, the partition function
diverges when $D\leq 4$.}
\label{sec:bosexact}

\subsection{Supersymmetric}
For $D=4$, $6$ and $10$, the Pfaffian is given
by \cite{Krauth:1998xh}
\beq
\C P=\left[ {8 \over 3} \Tr \comm{X_\mu}{X_\nu} \comm{X_\nu}{X_\rho}
\comm{X_\rho}{X_\mu} \right]^{D-2 \over 2}
=
\left[6\left( \lambda^2 V + \xi_i M_{ij} \xi_j \right) \right]^{D-2 \over 2},
\label{pfa}\eeq
where
\beq
M_{lm}=\delta_{lm} V - 2\left( v^2
u_l u_m -u\cdot v (v_l u_m + u_l v_m) +u^2 v_l v_m   \right).
\eeq
We note that $M$ and $U$ have the same eigenvectors. Specifically,
choosing $d-2$ linearly independent vectors $a$ each perpendicular
to $u$ and $v$, we have
\bea
U a &=& W a,\quad M a = V a\\
U u &=& v^2 u-u\cdot v v,\quad M u = 0 \\
U v &=& u^2 v-u\cdot v u,\quad M v = 0.
\eea
The eigenvalues of $U$ on the $u$-$v$ space are $q_{d-1}$ and $q_d$
where $q_{d-1}q_d= V$.


Proceeding as in the bosonic case and working in
the eigenvector basis of $U$ and
$M$ gives
\bea
L(p)&=& -{d^2 \over dp^2} \int d \lambda e^{ip\lambda/\sqrt{2}} d \xi du dv
\left( \lambda^2 V + V \sum_{k=1}^{d-2} \xi_k^2 \right)^{D-2 \over
2}\nonumber \\ & & \exp \left
( -\half W \sum_{k=1}^{d-2}\xi_k^2 -\half q_{d-1} \xi_{d-1}^2 - \half
q_d \xi_d^2 - \quarter V -\half \lambda^2 W  \right).
\eea
As before, we can use the rotation symmetry to do the $u$ and $v$
integrals;
the Pfaffian terms are pulled down by differentiating with respect to
$W$;  and then  $\xi_i$ is integrated out to give
\beq
L(p)={d^2 \over dp^2} \int d \lambda e^{ip\lambda/\sqrt{2}} \int_0^\infty dA
\int_0^{A^2/4}dB B^{D-3 \over 2} B^{D-4 \over 2} e^{-B/4} \left
( -{\partial \over \partial A}  \right)^{D-2 \over 2} A^{3-D \over 2}
e^{-\lambda^2 A/2}
\eeq
from which the eigenvalue density $\rho(\lambda)$ can be read off.
Integrating out $\lambda$ gives
\beq
L(p)= -\C N_D {d^2 \over d p^2} \int_0^\infty dA \,
A^{2D-6} e^{-A^2/16}\left( -{\partial \over \partial A} \right)^{D-4
\over 2} A^{2-D \over 2} e^{-{p^2 \over 4A}},
\eeq
where the overall constant factor
\beq
\C N_D={2^{3(3-D)}(D-3)(D-2)\sqrt{\pi}\over \Gamma\left( {D-3 \over 2 } \right)^2}
\eeq
is fixed by requiring $L(0)=1$. Then, by the saddle point method,
\beq
L(p) \sim (-1)^{{D\over 2}+1} 2^{3-{5D \over 6}} \sqrt{\pi \over 3} \C
N_D  \, p^{4D-12\over 3} \,\exp\left( - {3 \over 16} 2^{2 \over 3} \,
p^{4 \over 3}  \right)
\eeq
as $p \rightarrow \infty$.
Note that the asymptotic behaviour is exponential and that the power law
terms present in the bosonic case have disappeared. It is tempting to deduce
that this is a consequence of the supersymmetry. However it is easy to
 check that replacing the $\half(D-2)$ power in the expression for the
Pfaffian (\ref{pfa}) by any positive integer
 has the same effect of suppressing 
the power law terms but does not in general come from a supersymmetric
 theory\footnote{We thank Bergfinnur Durhuus for pointing this out to us.}.
The quantity appearing in (\ref{pfa}) is special in $su(2)$ because it is
proportional to a derivative of the action when expressed in appropriate
variables; it is the fact that one of the integrals is then 
of a total derivative
that suppresses the power law.

\section{Upper Bounds on Bosonic Large $\BS p$ Behaviour}
By using the convergence techniques of
\cite{Austing:2001bd,Austing:2001pk} we can obtain an 
upper bound on the large $p$ behaviour of the Polyakov line for
$su(N)$. We begin by illustrating the method on $su(2)$ and then on
$su(3)$ before giving the general argument.

\subsection{$\BS{su(2)}$}
Starting from  (\ref{Lsu2A}) 
we integrate out $\lambda_D$,  giving
\beq\label{Lsu2B} L(p)={\sqrt{\pi}\over 2}\int  
\prod_{k=1}^{D-1} dX_k \,
 Q^{-\threehalves}\left(2-\frac{p^2}{2Q}\right)
\exp\left(-\frac{p^2}{8 Q} -S_{D-1}\right),\eeq
where
\bea\label{Lsu2C} Q&=&\sum_{k=1}^{D-1}\absval{ \Xperp_k^1}^2,\nn\\
		   S_{D-1}&=&-\quarter \sum_{j,k=1}^{D-1}\Tr \comm{X_j}{X_k}^2,
\eea
so that a bound on $L(p)$ is given by
\beq
\absval{L(p)} < {\sqrt{\pi}\over 2}\int  
\prod_{k=1}^{D-1} dX_k \,
 Q^{-\threehalves}\left(2+\frac{p^2}{2Q}\right)
\exp\left(-\frac{p^2}{8 Q} -S_{D-1}\right).
\eeq
In the region of integration which leads to power law behaviour, we
must have
$Q \rightarrow \infty$ but $S_{D-1}$ bounded, so we will eventually need to
examine the flat directions of the $(D-1)$-dimensional action
$S_{D-1}$. First though, in any region in which $Q^{-1}>p^{-2+2\epsilon}$ with
$\epsilon$ a small positive parameter, we have exponential behaviour
of $L(p)$, so we restrict attention to the region
$Q^{-1}<p^{-2+2\epsilon}$, giving the bound
\beq
\absval{L(p)} < {\sqrt{\pi}\over 2}\int  
\prod_{k=1}^{D-1} dX_k \,
 p^{-3+3\epsilon}\left(2+\half p^{2\epsilon}\right)
\exp\left(-\frac{p^2}{8 Q} -S_{D-1}\right).
\eeq
In order to consider the flat directions, we use the radial variable $R$ defined by
\beq X_k=Rx_k,\quad \sum_{k=1}^{D-1}\Tr x_k x_k=1. \label{P4}\eeq
Then certainly $Q<R^2$, so we have the bound
\beq\label{2.19}
\absval{L(p)} < p^{-3+3\epsilon}\left(2+\half p^{2\epsilon}\right){\sqrt{\pi}\over 2}\int  
\prod_{k=1}^{D-1} dX_k \,
\exp\left(-\frac{p^2}{8 R^2} -R^4 \hat{S}_{D-1}\right)
\eeq
where
\beq
\hat{S}_{D-1}=-\quarter \sum_{j,k=1}^{D-1}\Tr \comm{x_j}{x_k}^2.
\eeq
To estimate the integral at large $p$ we
 split the integration domain into two:
 \bea {\C R_1}:&&  \hat S_{D-1} \geq R^{\epsilon-4}, \nn \\
 {\C R_2}:&&  \hat S_{D-1}  < R^{\epsilon-4}, 
\eea
where  $\epsilon$ is a small positive constant.
Since we know that the integrals are convergent the 
$R$ integral in  $\C R_1$ may be evaluated by the saddlepoint 
method and we get a contribution to $L(p)$ which 
decays exponentially at large $p$.
On the other hand the contribution from $\C R_2$
cannot be treated in this way and generates power law behaviour. 
Bounds on the angular integrals in (\ref{2.19}) have been
computed in \cite{Austing:2001bd} and  for $su(2)$, they give
\beq
\absval{L(p)} < p^{-3+3\epsilon}\left(2+\half p^{2\epsilon}\right){\sqrt{\pi}\over 2}\int  
dR \,
\exp\left(-\frac{p^2}{8 R^2} \right) 
R^{-D+4+\epsilon'},
\eeq
where $\epsilon'$ is another arbitrarily small positive parameter, so that
\beq
\absval{L(p)} < \const \,  p^{2-D + \epsilon''}
\eeq
as long as $D \geq 6$. 

The difficulty in finding a lower bound is illustrated in this
example. In (\ref{Lsu2B}), we have a positive term and a negative term,
and each gives the same upper bound power law behaviour. One can apply
the lower bound methods of \cite{Austing:2001pk}, and find the same power
law behaviour as lower bounds for each term (up to arbitrarily small
parameters $\epsilon$ again). These terms originate from the
Vandermonde factors and, since they each have the same
power law behaviour, we cannot rule out cancellation between them.

\subsection{$\BS {su(N>2)}$}
Completing the square in  (\ref{2.12}) and 
changing variables from $\lambda$ to $\nu= Q^{\half} \lambda$ gives,
\bea L(p)&=&
\int {d \underline{\nu} \over \det(Q)^\half} \prod_{k=1}^{D-1} dX_k \, \weyldet{\left(Q^{-\half}\nu+ip Q^{-1}\omega_1\right)}
\exp(-p^2 \omega_1\cdot Q^{-1}\cdot \omega_1)\nn\\
&&\times \exp \left(-\nu\cdot \nu+ \sum_{j,k=1}^{D-1}
\Tr \comm{X_j}{X_k}^2 \right),\label{Lnu}\eea
where
\bea
\label{qsu3}
Q_{ij}&=&\sum_{\alpha >0} {\alpha^i \alpha^j \over \absval{\alpha}^2}
T_\alpha\nn\\
T_\alpha &=& \sum_{k=1}^d \absval{X_k^\alpha}^2.
\eea
The integrand in (\ref{Lnu}) would be positive
definite but for the Weyl determinant factor which we may 
bound above
\bea  \absval{\weyldet{\left(Q^{-\half}\nu+ip Q^{-1}\omega_1\right)}} &<&
\const \! \left[  \Sumroot{\alpha} \!\! \left\{ (\alpha^T Q^{-\half} \nu)^2 + p^2
(\alpha^T Q^{-1} \omega_1)^2  \right\} \right]^{g-r \over 2}\nonumber \\
&=& \const \left[ \nu^T Q^{-1} \nu + p^2
(\omega_1^T Q^{-2} \omega_1)  \right]^{g-r \over 2},
\label{Weylbnd}
\eea
where we have used the relation
\beq
\Sumroot{\alpha} \alpha \alpha^T = \const \, 1.
\eeq

\subsubsection{$\BS{su(3)}$}
Now we will temporarily restrict the algebra to $su(3)$ and choose the  basis
\newline $s_1= (\sqrt{3 \over 2},-\sqrt{1\over 2})$, $s_2= (0,\sqrt{2})$,
in which, abreviating $T_i \equiv T_{\alpha_i}$, 
\beq
Q^{-1}= {1 \over 3(T_1T_2 +T_2T_3 + T_3T_1)}\left( 
\begin{array}{ccc}
 4T_1+T_2+T_3&  &-\sqrt{3} (T_2-T_3)\\
\\
-\sqrt{3} (T_2-T_3) & & 3(T_2+T_3)
\end{array}
\right)
\eeq
and
\beq\label{cruxsu3}
\omega_1^T Q^{-1} \omega_1 = {2 \over 9} {4T_1+T_2+T_3 \over  (T_1T_2 +T_2T_3 + T_3T_1)}.
\eeq
Since the $T_a$ are positive,  we can bound any matrix
element by
\beq\label{meltbnd}
x^T Q^{-1} y < \const \absval{x}\absval{y} \omega_1^T Q^{-1} \omega_1,
\eeq
which applies in particular to the eigenvalues of $Q^{-1}$. Taking these bounds together with
(\ref{Weylbnd}), we can bound the expression (\ref{Lnu}) for $L(p)$
\bea \absval{L(p)}&<&
\int d \underline{\nu} (\omega_1^T Q^{-1} \omega_1)^{r/2}
\prod_{k=1}^{D-1} dX_k \, (\omega_1^T Q^{-1} \omega_1)^{(g-r)/2} 
\sum_{n=0}^{(g-r)/2} a_n (\nu \cdot \nu)^{(g-r)/2-n}
\nonumber\\&& (p^2 \omega_1^T Q^{-1} \omega_1)^{n}
\exp(-p^2 \omega_1^T Q^{-1} \omega_1)\nn \exp \left(-\nu\cdot \nu+ \sum_{j,k=1}^{D-1}
\Tr \comm{X_j}{X_k}^2 \right),\\\eea
where the $a_n$ are the coefficients of the binomial expansion. We can
now integrate out $\nu$ giving some new positive coefficients
$\tilde{a}_n$, 
\beq \label{b}
\absval{L(p)}<\int  
\prod_{k=1}^{D-1} dX_k \,
\sum_{n=0}^{(g-r)/2}\tilde{a}_n p^{2n}(\omega_1^T Q^{-1}
 \omega_1)^{n+g/2}  
\exp\left(-p^2 \omega_1^T Q^{-1} \omega_1 -S_{D-1}(X) \right).\eeq
We now split the integration region into two
parts. In the region in which $\omega_1^T Q^{-1} \omega_1 \geq p^{-2+2
\epsilon}$ with $\epsilon$ small but positive, we have exponential
behaviour in $p$. We therefore restrict our attention to the region in which
\beq\label{b1}
\omega_1^T Q^{-1} \omega_1 < p^{-2+2 \epsilon}.
\eeq
Writing $R$ as the
radial variable $R^2 = \widetilde{X}_\mu^i  \widetilde{X}_\mu^i + \overline{X}_\mu^\alpha
\overline{X}_\mu^{-\alpha}$ (with summation over $\mu$, $i$ and $\alpha$ implied)
we have the bound
\beq\label{b2}
\omega_1^T Q^{-1} \omega_1 > {A \over R^2},
\eeq
where $A$ is a constant, since the $T_a$ are positive in (\ref{cruxsu3}).
 Inserting (\ref{b1}) and
(\ref{b2}) into (\ref{b}) gives
\beq\label{b3} 
\absval{L(p)}<
\sum_{n=0}^{(g-r)/2}\tilde{a}_n p^{2n}(p^{-2+2 \epsilon})^{n+g/2}\int  
\prod_{k=1}^{D-1} dX_k \,  
\exp \left(-A{p^2 \over R^2} \right)\nn \exp \left( -S_{D-1}(X) \right).
\eeq
If $0 <R<1$, the behaviour is exponential in $p$. For $R>1$, a bound on the angular part of the integral has been calculated in
\cite{Austing:2001bd}, and we can insert the result directly into (\ref{b3})
to obtain
\beq
\absval{L(p)}<
\sum_{n=0}^{(g-r)/2}\tilde{a}_n p^{-g+\epsilon (2n+g)} \int_1^\infty dR \exp\left
( -A {p^2 \over R^2}\right)R^{-k_c(D-1)-1 +\epsilon'},
\eeq
where $\epsilon'$ is another arbitrarily small positive number, and
\beq
k_c(D)=2rD-D-4r-\delta_{D,3}\delta_{r,2}\label{kceqn}.
\eeq
Finally then, we obtain
\beq
\absval{L(p)}<\const \; p^{-k_c(D-1)-g +\epsilon''},
\eeq
where $\epsilon''$ is a new arbitrarily small but positive number.

These arguments can be extended to some correlation functions of
Polyakov lines. Consider
\bea L(p,q)&=&<\Tr\exp(i p X_1 ) \Tr\exp(i q X_1 )>\nn\\
&=& N \int d \lambda_D  \Delta^2_G(\lambda_D) dX_k
\, \sum_h\exp(i p \lambda_D\cdot h )
\, \sum_{h'}\exp(i q \lambda_D\cdot h' )\nn\\
&&\quad\times\exp \left(-\sum_{\alpha >0}
\frac{(\lambda_D \cdot \alpha)^2}{\absval{\alpha}^2}
\sum_{k=1}^{D-1}
\absval{ \Xperp_k^\alpha}^2-S_{D-1} \right).
\eea
Clearly the terms with $h=h'$ satisfy the same bound as
$L(p)$ with $p$ replaced by $p+q$. After completing the square,
 the cross terms become 
\bea L(p)&=&
\int {d \underline{\nu} \over \det(Q)^\half} \prod_{k=1}^{D-1} dX_k \, \weyldet{\left(Q^{-\half}\nu+i Q^{-1}\sigma\right)}
\exp(- \sigma\cdot Q^{-1}\cdot \sigma)\nn\\
&&\quad\times\exp \left(-\nu\cdot \nu+ \sum_{j,k=1}^{D-1}
\Tr \comm{X_j}{X_k}^2 \right),\label{LLnu}\eea
where, for example (and specializing to
$su(3)$),
\beq \sigma=p\omega_1-q\omega_2.\eeq
Introducing an arbitrary constant $0<\beta<1$ we find that 
\bea \sigma^T Q^{-1} \sigma &=& {4 \over 9 (T_1T_2 +T_2T_3 + T_3T_1)}\,
\Big( \beta \vert \sigma\vert^2 (T_1+T_2+T_3)
+T_1((2p-q)^2-\beta\vert \sigma\vert^2)\nn\\
&&\quad +T_3((p-2q)^2-\beta\vert \sigma\vert^2)
+T_2((p+q)^2-\beta\vert \sigma\vert^2)\Big).
\eea
It follows that, in the region
\bea (2p-q)^2-\beta\vert \sigma\vert^2=(2p-q)^2-\beta(p^2+pq+q^2)>0, \nn\\
(2q-p)^2-\beta\vert \sigma\vert^2=(2q-p)^2-\beta(p^2+pq+q^2)>0,\label{cone}
\eea
we have
\beq \sigma^T Q^{-1} \sigma \ge \beta \vert \sigma\vert^2 (\mu_1+\mu_2)\ge const\frac{ \beta\vert \sigma\vert^2 }{R^2},\eeq
where $\mu_{1,2}$ are the eigenvalues of $Q^{-1}$ so that
\beq \absval{ x^T  Q^{-1} y} \le \absval x \absval y\frac{ \sigma^T Q^{-1} \sigma}
{\beta \absval \sigma ^2}\eeq
and
\beq \label{bb}
\absval{L(p)}<
\sum_{n=0}^{(g-r)/2}\tilde{a}_n' \absval \sigma ^{-g}(\sigma^T Q^{-1} \sigma)^{n+g/2}\int  
\prod_{k=1}^{D-1} dX_k \,  
\exp(-\sigma^T Q^{-1} \sigma)\nn \exp \left( -S_{D-1}(X) \right),\eeq
where the constants $\tilde{a}_n'$ depend on $\beta$. Again, unless
$\sigma^T Q^{-1} \sigma< \absval \sigma ^{2\epsilon}$ we get exponential behaviour
so  proceeding as for $L(p)$ we obtain
\beq L(p,q)< const \absval \sigma ^{-k_c(D-1)-g +\epsilon'''}.\label{ubound}\eeq
Keeping $\beta$ small but fixed, we see that in the regions not satisfying
(\ref{cone}) $\sigma$ is almost proportional to a root. Since the roots are
 weights
for the adjoint representation there is screening and it is easy to see
that we just get a constant upper bound instead of (\ref{ubound}).

\subsubsection{$\BS{su(N>3)}$}
We now need to
 generalise the formula \ref{cruxsu3} for $\omega^T Q^{-1} \omega$
 to $SU(N)$.
First we show that for $Q$ defined as in (\ref{qsu3}),
\beq\label{detq}
\det Q = {r+1 \over 2^r} \sum_{ \{ \alpha_1 , \cdots , \alpha_r \}
\in \C S(r)} T_{\alpha_1} \cdots T_{\alpha_r},
\eeq
where $\C S(n)$ is the set of all n-tuples of positive roots $\alpha$ which are
linearly independent.
To see this, we use the fact that the determinant is a homogeneous
function of the $T_\alpha$ of degree $r$. To find the coefficient of
$T_{\alpha_1} \cdots T_{\alpha_r}$ for any choice of the $\{ \alpha_1
\cdots \alpha_r \}$, set $T_{\alpha_i}=1$, $i=1, \cdots , r$ and
$T_\alpha=0$ otherwise. Then $Q$ becomes
\beq
Q_{ij}=\half \sum_{k=1}^r{\alpha_k^i \alpha_k^j }
\eeq
and
\beq \det Q=2^{-r} \det ({\alpha}_1, \cdots, {\alpha}_r)
\det ({\alpha}_1, \cdots, {\alpha}_r)^T,
\eeq
 where
$({\alpha}_1, \cdots, {\alpha}_r)$ is the matrix whose columns are the
${\alpha}_i$. If $\{ \alpha_1,
\cdots, \alpha_r \}$ are linearly dependent then $\det Q=0$. If $\{ \alpha_1,
\cdots, \alpha_r \}$ are linearly independent, then,
since for $su(N)$ any root 
 is a Weyl transformation of any other root and
any positive root can be written in terms of simple roots as
\beq
\label{rootsun}
\alpha= s_j + s_{j+1} + \cdots + s_{k-1}+s_k
\eeq
with $1 \leq j \leq k \leq r$, we can use a series
of Weyl transformations and column operations to reduce 
\beq\label{alphareduce}
({\alpha}_1, \cdots, {\alpha}_r) \rightarrow (\pm {s}_1, \cdots, \pm {s}_r).
\eeq
Any minus signs can be dropped since we are only after 
the square of the determinant. The product of 
$({s}_1, \cdots, {s}_r)$ and its transpose is just the Cartan
matrix whose determinant is $r+1$
so
\beq
\det Q ={r+1 \over 2^r},
\eeq
which completes the proof of (\ref{detq}).
\label{ap1}

Now we show  that
\beq\label{crux1}\label{crux}
\omega_1^T Q^{-1} \omega_1 = 2{\sum_{ s=\{ \alpha_1 , \cdots , \alpha_{r-1} \}
\in \C S(r-1)} \left( {n_s \over N} \right)^2 T_{\alpha_1} \cdots T_{\alpha_{r-1}} \over  \sum_{ \{ \alpha_1 , \cdots , \alpha_r \}
\in \C S(r)} T_{\alpha_1} \cdots T_{\alpha_r}},
\eeq
where the $n_s$ are non-zero integers. This will allow us to use the
same form of bound (\ref{meltbnd}) as for $SU(3)$.

It is helpful to use a basis in which 
\beq
\omega_1= \sqrt{r \over r+1}\left(
\begin{array}{c}
1\\
0\\
\vdots\\
0
\end{array}
\right).
\eeq
In this basis
\beq\omega_1^T Q^{-1} \omega_1= {r \over r+1}
\Qperp_{1,1}/\det Q,\eeq
 where $\Qperp$ is the
matrix of cofactors. 
 Our task then is to compute the cofactor
$\Qperp_{1,1}$, and as this is a homogeneous function of the $T_\alpha$
of degree $r-1$, we can set $T_{\beta_{(1)}}= \cdots = T_{\beta_{(r-1)}}=1$
and $T_\alpha=0$ otherwise, as before.
Then 
\beq\Qperp_{1,1}=2^{1-r}\det \Qbar,\eeq
 where the matrix $\Qbar$ is given by
\beq
\Qbar= \left(
\begin{array}{cccc}
1&0& \cdots& 0\\
0\\
\vdots && \sum_{n=1}^{r-1} \tilde{\beta}_{(n)} \tilde{\beta}_{(n)}^T \\
0
\end{array}
\right)
\eeq
with
\beq
\begin{array}{l}
\tilde{\beta}_{(n)}^1=0\\
\tilde{\beta}_{(n)}^i={\beta}_{(n)}^i \, \;\;\; i \neq 1.
\end{array}
\eeq
Proceeding as before we note that 
\beq
\det \Qbar = \det \left(
\begin{array}{cccc}
1&0&\cdots&0\\
0\\
\vdots&\tilde{\beta}_{(1)}&\cdots&\tilde{\beta}_{(r-1)}\\
0
\end{array}
\right)^2.
\eeq
So, using linear column operations, we can restore the $\tilde{\beta}_{(i)}$
to ${\beta}_{(i)}$ to get
\beq
\det \left(
\begin{array}{cccc}
1&0&\cdots&0\\
0\\
\vdots&\tilde{\beta}_{(1)}&\cdots&\tilde{\beta}_{(r-1)}\\
0
\end{array}
\right)
=\sqrt{r+1 \over r}\det (\omega_1, {\beta}_{(1)}, \cdots, \beta_{(r-1)}).
\eeq
If the $\beta_{(i)}$ are linearly dependent, then $\det
\Qbar=0$. Using the same procedure as before, we obtain
\beq
\det (\omega_1, {\beta}_{(1)}, \cdots, {\beta}_{(r-1)}) = \det (w \omega_1, s_{(1)}, \cdots, s_{(r-1)}),
\eeq
where $w$ is a Weyl transformation, and the $s_{(i)}$ are $r-1$ of
the simple roots. We note that this determinant can be zero in some
representations of the algebra. For example, in the adjoint
representation in which the weights are roots. However, we now show that it is
never zero in the fundamental representation. First, since any Weyl
transformation is the product of Weyl reflections in roots, we have
\beq
w \omega_1 = \omega_1 - 2 \sum_{i=1}^r p_i {s}_i
\eeq
where the $p_i$ are integers. In terms of the simple roots, $\omega_1$ is given by
\beq
\omega_1 ={1 \over r+1} \left[ r {s}_1 +
(r-1){s}_2 + \cdots +{s}_r \right]
\eeq
so we find
\bea\label{2.60}
\det  (w \omega_1, s_{(1)}, \cdots, s_{(r-1)})& =& \det  ({q \over
r+1} s_{(r)} - 2 n s_{(r)}, s_{(1)}, \cdots, s_{(r-1)})\nn\\
&=&\pm {n_s \over r+1} \det (s_1, \cdots, s_r),
\eea
where $q$ is an integer between $1$ and $r$,  $n$ is one of the
$p_i$, also an integer, and $n_s$ is a non-zero integer.
 Then we have
\beq
 \det \Qbar = \frac{n_s^2}{r}
\eeq
and, together with the expression (\ref{detq}) for $\det Q$, this completes the result (\ref{crux}).
\label{ap2}

The upper bound for $SU(N)$ now follows in
 a manner identical to the proof of (\ref{crux1}). The cofactors of $Q$ take the
general form
\beq
Q^\perp_{ij}= \sum_{s=(\alpha_1 \cdots \alpha_{r-1}) \in\C S(r-1)} C_s \,T_{\alpha_1}
\cdots T_{\alpha_{r-1}},
\eeq
where the $C_s$ are positive constants, so that we have again the result (\ref{meltbnd})
\beq
x^T Q^{-1} y < \const\, \absval{x}\absval{y} \omega^T Q^{-1} \omega.
\eeq
The remainder of the argument follows without modification from the
$su(3)$ case, and we have again
\beq
\absval{L(p)}< \const \, p^{-k_c(D-1)-g+\epsilon},\label{suNupper}
\eeq
where the function $k_c$ is given in \ref{kceqn}, and
$\epsilon$ is arbitrarily small.

\section{Conclusions}
For the bosonic $su(2)$ model the exact calculation shows that asymptotically
$L(p)\sim p^{2-D}$ while the upper bound (\ref{su2upper})
 demonstrates that this behaviour  arises from the flat directions in the
action. For larger gauge groups there is unfortunately no exact calculation
 but again the flat directions lead to a  power law bound on $L(p)$ at large
 $p$
(\ref{suNupper}). In the process of proving the bound we throw away the
 alternating signs coming from the Vandermonde in (\ref{Lnu}); it is possible 
that in an exact calculation these would lead to the cancellation of the 
leading power of $p$. However, as we remarked previously, this does not
 happen in $su(2)$ and so there is no reason to suppose that it happens for
 larger groups. We conclude that $L(p)$ has the asymptotic behaviour
\beq {L(p)}\sim \const \, p^{-k_c(D-1)-g}
\eeq
at large $p$. Our proof of this bound uses certain properties that are
 special to the $su(N)$ Lie algebra but, just as the convergence proofs
 go through for all gauge groups, we expect that very similar results 
for $L(p)$ will hold for
all the compact Lie groups.  This power law property is different from
 the exponential asymptotic behaviour given by the mean field approximation
 \cite{Oda:2000im}.
 The reason is that the 
mean-field saddle point computation misses the dominant region of
 integration at large $p$ which is essentially the end-point of the
 integration domain rather than the saddle point.

Although the asymptotic behaviour of the supersymmetric $su(2)$ model is 
qualitatively different, being exponential rather than power law, it 
is not clear whether this extends to supersymmetric $su(N)$.
 In the $su(2)$ model the
power law is suppressed whenever there are fermions in the adjoint
 representation so this is a feature of $su(2)$ rather than of supersymmetry.
The asymptotic behaviour of the  supersymmetric $su(N)$ models is an open question.

\acknowledgments{We are grateful to Bergfinnur Durhuus and
 Thordur Jonsson for useful discussions. PA acknowledges an EPSRC fellowship.
GV acknowledges the support of the European network on ``Discrete
 Random Geometries'' HPRN-CT-1999-00161 EURO-GRID.
}
\providecommand{\href}[2]{#2}\begingroup\raggedright\endgroup

\end{document}